\documentstyle [12pt,twoside]{article}

\newcommand{\bq}{\begin{equation}}
\newcommand{\ba}{\begin{eqnarray}}
\newcommand{\eq}{\end{equation}}
\newcommand{\ea}{\end{eqnarray}}

\def\d{\delta}

\def\g{\raisebox{.4ex}{$\gamma$}}

\def\l{\lambda}

\def\G{\Gamma}

\def\bo{{\raise.15ex\hbox{\large$\Box$}}}
\def\bob{{\lower.2ex\hbox{\large$\Box$}}}
\def\pa{\partial}

\def\TH{{\raise.2ex\hbox{$\displaystyle \bigodot$}\mskip-4.7mu \llap H \;}}
\def\face{{\raise.2ex\hbox{$\displaystyle \bigodot$}\mskip-2.2mu \llap {$\ddot
        \smile$}}}

\def\Hat#1{\rlap{\kern.10em$\widehat{\phantom G}$}#1}
\def\HAt#1{\rlap{\kern.05em$\widehat{\phantom G}$}#1}

\def\cap#1{\rlap{\kern.1em$\widehat{\phantom{G\vrule height.8em}}$}#1{}}
\def\Cap#1{\rlap{\kern.05em$\widehat{\phantom{G\vrule height.8em}}$}#1{}}

\def\VEV#1{\left\langle #1\right\rangle}

\def\abs#1{\left| #1\right|}
\def\leftrightarrowfill{$\mathsurround=0pt \mathord\leftarrow \mkern-6mu
        \cleaders\hbox{$\mkern-2mu \mathord- \mkern-2mu$}\hfill
        \mkern-6mu \mathord\rightarrow$}
\def\overleftrightarrow#1{\vbox{\ialign{##\crcr
        \leftrightarrowfill\crcr\noalign{\kern-1pt\nointerlineskip}
        $\hfil\displaystyle{#1}\hfil$\crcr}}}

\def\frac#1#2{{\textstyle{#1\over\vphantom2\smash{\raise.20ex
        \hbox{$\scriptstyle{#2}$}}}}}

\def\afrac#1#2{{\vphantom1\smash{\lower.5ex\hbox{$#1$}}\over#2}}

\catcode`@=11
\def\underline#1{\relax\ifmmode\@@underline#1\else
        $\@@underline{\hbox{#1}}$\relax\fi}
\catcode`@=12

\def\nis{\nointerlineskip}
\def\Abar{\vbox{\nis\moveright.33em\vbox{
        \hrule width.35em height.04em}\nis\kern.05em\hbox{$A$}}{}}
\def\Dbar{\vbox{\nis\moveright.20em\vbox{
        \hrule width.50em height.04em}\nis\kern.05em\hbox{$D$}}{}}
\def\Gbar{\vbox{\nis\moveright.20em\vbox{
        \hrule width.50em height.04em}\nis\kern.05em\hbox{$G$}}{}}
\def\mbar{\vbox{\nis\moveright.15em\vbox{
        \hrule width.60em height.04em}\nis\kern.05em\hbox{$m$}}{}}
\def\Rbar{\vbox{\nis\moveright.20em\vbox{
        \hrule width.50em height.04em}\nis\kern.05em\hbox{$R$}}{}}
\def\Vbar{\vbox{\nis\moveright.05em\vbox{
        \hrule width.60em height.04em}\nis\kern.05em\hbox{$V$}}{}}
\def\Xbar{\vbox{\nis\moveright.20em\vbox{
        \hrule width.60em height.04em}\nis\kern.05em\hbox{$X$}}{}}
\def\thetabar{\vbox{\nis\moveright.15em\vbox{
        \hrule width.30em height.04em}\nis\kern.05em\hbox{$\theta$}}{}}
\def\Lambdabar{\vbox{\nis\moveright.25em\vbox{
        \hrule width.35em height.04em}\nis\kern.05em\hbox{${\mit\Lambda}$}}{}}
\def\Sigmabar{\vbox{\nis\moveright.25em\vbox{
        \hrule width.50em height.04em}\nis\kern.05em\hbox{${\mit\Sigma}$}}{}}
\def\phibar{\vbox{\nis\moveright.18em\vbox{
        \hrule width.40em height.04em}\nis\kern.05em\hbox{$\phi$}}{}}
\def\chibar{\vbox{\nis\moveright.12em\vbox{
        \hrule width.40em height.04em}\nis\kern.05em\hbox{$\chi$}}{}}
\def\psibar{\vbox{\nis\moveright.23em\vbox{
        \hrule \def\debar{\vbox{\nis\moveright.18em\vbox{
        \hrule width.35em height.04em}\nis\kern.05em\hbox{$\partial$}}{}}}}}
\def\delbar{\vbox{\nis\moveright.10em\vbox{
        \hrule width.63em height.04em}\nis\kern.05em\hbox{$\nabla$}}{}}

\newskip\humongous \humongous=0pt plus 1000pt minus 1000pt

\newif\ifdtup

\def\begintitle#1#2#3#4
        {\begin{titlepage}
         \centerline{#1 \hfill MdDP-PP-88-#2}
         \begin{center}\vglue .7in
         {\large\bf #3}\\].7in(
         {\bf #4}\\
         {\it Department of Physics and Astronomy}\\
         {\it University of Maryland, College Park, MD 20742}\\].7in(
         {\bf ABSTRACT}\\
         \end{center}
         \begin{quotation}}
\def\endtitle
         {\end{quotation}
          \end{titlepage}
          \newpage}

\begin{document}
\centerline{\large{\bf Nonlinear Noise in Cosmology}}
\vskip 1.5cm
\centerline{Salman Habib$^{\star}$ and Henry E. Kandrup$^{\dagger}$}
\vskip .5cm
\centerline{\em $^{\star}$T-6, Theoretical Astrophysics}
\centerline{\em Los Alamos National Laboratory}
\centerline{\em Los Alamos, NM  87545}
\vskip .5cm
\centerline{\em $^{\dagger}$Department of Astronomy and Institute for
Fundamental Theory}
\centerline{\em University of Florida}
\centerline{\em Gainesville, FL 32611}
\vskip .5in
\centerline{\bf Abstract}
\vskip .1in
{\small
This paper derives and analyzes exact, nonlocal Langevin equations
appropriate in a cosmological setting to describe the interaction of
some collective degree of freedom with a surrounding ``environment.''
Formally, these equations are much more general, involving as they do
a more or less arbitrary ``system,'' characterized by some
time-dependent potential, which is coupled via a nonlinear,
time-dependent interaction to a ``bath'' of oscillators with
time-dependent frequencies. The analysis reveals that, even in a
Markov limit, which can often be justified, the time dependences and
nonlinearities can induce new and potentially significant effects,
such as systematic and stochastic mass renormalizations and
state-dependent ``memory'' functions, aside from the standard
``friction'' of a heuristic Langevin description. One specific example
is discussed in detail, namely the case of an inflaton field,
characterized by a Landau-Ginsburg potential, that is coupled
quadratically to a bath of scalar ``radiation.'' The principal
conclusion derived from this example is that nonlinearities and
time-dependent couplings do {\em not} preclude the possibility of
deriving a fluctuation-dissipation theorem, and do {\em not} change
the form of the late-time steady state solution for the system, but
{\em can} significantly shorten the time scale for the approach
towards the steady state.}
\vfill
\noindent e-mail:\\
\noindent habib@eagle.lanl.gov\\
\noindent kandrup@astro.ufl.edu

\newpage

\centerline{\bf I. Introduction}
\vskip .1in
Over the past two decades or so, a great deal of attention has focused
on the problem of understanding statistical field theory in a
cosmological context, allowing correctly for the fact that the
spacetime is not flat, or even static. This is a problem of interest
both from an abstract field theoretic viewpoint and from a more
practical viewpoint which seeks to understand what the Universe was
like at very early times. When addressing the latter sorts of issues,
one is led oftentimes to implement various models and/or
approximations which, although not justified rigorously, seem
physically well motivated and even essential, if one wishes to obtain
concrete results.

One important ingredient in this sort of modeling is the idea that,
at some level, the physical degrees of freedom of the Universe divide
into two coupled pieces, a ``system'' component, the detailed
evolution of which is for some reason of particular relevance, and a
``bath'' component, the detailed evolution of which is somehow
irrelevant. This sort of picture has arisen in at least four different
settings. One such setting entails an understanding of inflation in
terms of an inflaton field evolving in the rest of the Universe, which
serves as an external environment or bath \cite{kn:rba}-\cite{kn:hka}.
Another involves the general notion of ``coarse-graining'' as a physical
mechanism in terms of which to extract quantum decoherence, this
facilitating a ``quantum-to-classical'' transition in the early
Universe \cite{kn:deco}. A third entails a more systematic development
of statistical quantum field theory \cite{kn:blha}, which uses a
closed-time-path formalism to derive quantum dissipation and memory
loss. And finally, there is the intriguing, but not yet completely
understood, program of stochastic inflation originally proposed by
Starobinsky \cite{kn:sta}.

Much work along these lines has been predicated upon the formulation
of essentially {\em ad hoc} Fokker-Planck equations, which effectively
introduce a bilinear interaction between the system and the bath. It
seems crucial to understand the extent to which these sorts of
heuristic models are in fact reasonable, i.e., approximately true in
some appropriate limit, and, especially, how relaxing the basic
assumption of bilinearity changes the underlying physics. In
particular, is the standard sort of modeling legitimate if one allows,
as in certain cases one must, for time-dependent couplings and
frequencies and incorporates realistic nonlinearities?

These are extremely difficult questions to answer in complete
generality. However, as will be seen in this paper, they {\em can} be
examined in certain cases by considering special models of systems
coupled to baths comprised of time-dependent harmonic oscillators, where
it is possible to derive exact, nonlocal Langevin equations simple enough
to understand, both physically and mathematically. These models are
nonlinear, time-dependent generalizations of phenomenological models
which have proven quite successful in other branches of physics, such
as condensed matter physics or quantum optics. And, as in those
settings, the models {\em are} well motivated phenomenologically, even
if they are not derived {\em ab initio}.

One of the objectives here is to derive exact generalized Langevin
equations for these models, and in an appropriate Markov limit, their
Fokker-Planck realizations. One is then poised to understand the sorts
of new effects arising in these exact nonlocal equations which are
absent both from more heuristic Langevin descriptions and from the
exact nonlocal equations that can be derived for the special case of
time-independent, bilinear couplings. What this entails is an analysis
of the interplay between three different sorts of effects, namely
linear and nonlinear noise, which may well have very different natural
timescales, and the explicit time-dependence of the environment,
reflecting the overall expansion of the Universe, which introduces yet
another timescale.

This analysis shows that (a) allowing for a nontrivial time-dependence
necessarily induces qualitatively new effects like a mass (or
frequency) ``renormalization,'' even for the special case of bilinear
couplings \cite{kn:hka}); and, moreover, (b) allowing for
nonlinearities in the system-environment coupling induces new effects
aside from the usual ``friction'' term. One discovers, e.g., that
nonlinearities give rise to an additional renormalization of the
system potential, and that they imply a ``memory'' kernel which
involves the state of the system.

These results might suggest that, in the presence of such
nonlinearities, one's naive intuition is completely lost. This,
however, is not so: Even allowing for nonlinearities and
time-dependent couplings in the interaction between the system and
environment, one can, at least for the case of time-independent
oscillators, where the bath may still be viewed as being ``at
equilibrium,'' still derive a simple fluctuation-dissipation theorem
\cite{kn:rz}. When the oscillators become time-dependent, the
fluctuation-dissipation theorem will no longer be exact. However, this
theorem {\em does} remain at least approximately true to the extent
that the coupling between the system and environment is dominated by
modes of sufficiently short wavelength. As will be shown below, in a
cosmological setting this implies that, on scales short compared with
the horizon length, it is still possible to speak of an approximate
equilibrium and an approach towards that equilibrium.

This paper focuses on obtaining a qualitative understanding of the
effects of nonlinear couplings and a time-dependent environment. A
subsequent paper will present a concrete calculation, applying the
technology of the time-dependent renormalization group to a simple
cosmological phase transition.

Section II of this paper focuses on the general problem of couplings
between a system and a bath, motivating in particular a rather general
time-dependent Hamiltonian which is amenable to a systematic analysis.
Section III derives an exact, nonlocal Langevin equation from this
Hamiltonian and then discusses its physical implications. All of this
is completely general, not restricted in any way to a cosmological
context. Section IV then turns to a consideration of one specific
cosmological model, deriving Langevin/Fokker-Planck equations for
some collective degree of freedom, such as the dilaton mode, evolving
in a Landau-Ginsburg potential and coupled to scalar ``radiation.''
This equation can provide one with a simple tool in terms of which to
model a cosmological phase transition associated either with inflation
or the formation of a cosmic string. Section V provides approximate
solutions to this equation, which enable one to study the approach
towards a (time-dependent) steady state. It is observed that such an
approach towards ``equilibrium'' can be strongly influenced by the
nonlinearities in the couplings (``multiplicative noise'') and/or the
effects of the time-dependent expansion of the Universe.
\newpage

\centerline{\bf II. The System-Environment Splitting}
\vskip .1in
Two sorts of equations are ubiquitous in nonequilibrium statistical
mechanics, namely (collisional) Boltzmann equations and
Fokker-Planck equations. Boltzmann equations are appropriate for
transport-type problems, involving strongly interacting particles,
whereas Langevin/Fokker-Planck equations are useful in the study of
phase transitions, Brownian motion, and the like.

These two ubiquitous equations also find a place in cosmology. In the
past, kinetic theory {\em a la} Boltzmann has been studied
extensively, most prominently perhaps in the analysis of
nucleosynthesis \cite{kn:nuc}. More recently, however, phase
transitions have come into vogue, especially with regard to the
inflationary scenario \cite{kn:inf}, and it is here that one
encounters Fokker-Planck equations.

That the standard methodology (with essentially trivial modifications)
can be applied to the very early Universe involves a certain leap of
faith; and while on the whole cosmologists appear comfortable with the
{\em status quo,} doubts have certainly been voiced in the literature
\cite{kn:md}. The main objections relate to (1) the assumption of
thermal equilibrium, (2) the absence of a clear separation of time
scales, (3) the neglect of fluctuations,  (4) the validity of
heuristic master equations, and, related to this, (5) the lack of a
fundamental Liouville description, derived from a Hamiltonian. While
it is fair to say that some of these objections have not been stated
in a concrete, quantitative way, they {\em do} lead to a feeling of unease.

In order to address some of these issues concretely, this paper will
consider a prototypical Hamiltonian which incorporates more or less
realistic nonlinearities and time dependences, and then extract
from that Hamiltonian an {\em exact} Langevin equation for the system
variable. This is a Liouville approach, \cite{kn:md}, in which the
fundamental equation is derived systematically without any {\em ad
hoc} assumptions. However, the more difficult problem of justifying a
full-blown nonlinear Boltzmann equation will not be treated here.

The exact Langevin equation involves at least three distinct time scales:
(1) the Hubble time $t_{H}$, (2) the relaxation time $t_{R}$ on which the
system is affected by the surrounding environment, and (3) the time scale
$t_{C}$ set by the decay of the noise autocorrelation function. If the
system evolves under the influence of some nontrivial system potential
$U$, there is also a fourth time scale $t_{S}$, the time scale on
which the system changes in response to $U$. The implementation of any
approximation entails an assumption regarding the separation of these
time scales. In particular, the possibility of an approximate local
description, i.e., the existence of a Markov limit, depends critically
on the assumption that $t_{C}{\;}{\ll}{\;}t_{R}$, $t_{H}$, and
$t_{S}$. An obvious point then is that nonlinearities, especially with
respect to the system-bath coupling, can play an important role by
inducing multiplicative noise which can significantly alter $t_{R}$.

The specific objective here is to motivate a phenomenological
Hamiltonian for the system plus environment, and to analyze it
rigorously to extract the physical effects contained therein. The
system is taken to be some ``collective coordinate'' evolving under
the influence of a ``heat bath.'' Particular interest focuses on phase
transitions, but the development here is in no way restricted to such
a setting. Indeed, it should be stressed that this general approach
has been used successfully in many other areas of physics, such as
condensed matter physics, nonlinear optics, nuclear physics, etc.

Given an arbitrary composite Hamiltonian, one wishes to introduce a
splitting into a ``system'' piece, a ``bath'' piece, and an interaction
term. If this split is to be useful, it must be true that in some
sense the system is ``small'' compared with the environment. What this
means is that, as far as the system is concerned, the full Hamiltonian
\bq
H_T=H_S+H_B+H_I \label{1}
\eq
is well approximated by
\bq
H=H_S+\delta H_B+\delta H_I \label{2}
\eq
where $\delta H_B$ is the Hamiltonian for a collection of harmonic
oscillators and $\delta H_I$ an interaction Hamiltonian linear in the
oscillator variables $q_A$ \cite{kn:nucref} \cite{kn:cl}. The heat
bath may well be one in which the oscillators are ``fundamental''
(e.g., the modes of some free field, as in black body photons), but
this is by no means necessary: Assume that, in the absence of any
coupling with the system, the environment is characterized by some
fixed, possibly time-dependent, solution. Now allow for a weak
coupling with the system, weak in the sense that each bath mode is
only changed marginally. Then identify the $q_A$'s as perturbed
variables, i.e., degrees of freedom defined relative to the fixed
solution (e.g., phonons). This has two implications: (1) the
environment can be visualized as a collection of oscillators with
(possibly time-dependent) frequencies, so that ${\delta}H_{B}$ is
quadratic in bath variables $q_A$, and, (2) because the interaction of
the environment with the system is assumed to be weak, in the sense
that the individual bath modes are not altered significantly,
${\delta}H_{I}$  must be linear in the $q_{A}$'s. Note that one does
{\em not} want to assume that the system is only weakly altered, so
the interaction ${\delta}H_{I}$ is {\em not} necessarily linear in the
system variable $x$. In principle one can proceed without imposing any
restrictions on the form of the system Hamiltonian $H_{S}$.

Given the above set of assumptions, one can write that
\bq
H_{S}={1\over 2}{\;}v^{2}+V_{ren}(x,t), \label{3}
\eq
\bq
\delta H_{B}={1\over
2}\sum_{A}\left[p_{A}^{~2}+{\Omega}_{A}^{~2}(t)q_{A}^{~2}\right],
\label{4}
\eq
and, in terms of relatively arbitrary functions $\Gamma_A$,
\bq
\delta H_{I}=-{\,} \sum_{A}{\,} \Omega_A^{~2}(t)\Gamma_{A}(x,t)q_{A}.
\label{5}
\eq
It is, however, convenient to rewrite $H$ in the manifestly positive
form
\bq
H={1\over 2}v^{2}+U(x,t) +
{1\over 2}\sum_{A}\left\{p_{A}^{~2}+{\Omega}_{A}^{~2}(t)
\left[q_{A}-{\Gamma}_{A}(x,t)\right]^2\right\}, \label{6}
\eq
where in terms of the ``renormalized'' potential $V_{ren}$,
\bq
U(x,t)=V_{ren}-{1\over 2}\sum_A\Omega_A^{~2}(t)\Gamma_A^{~2}(x,t).
\label{7}
\eq

Couplings of a system to some environment can induce finite and
stochastic renormalizations in the system potential, although this is
not always so \cite{kn:cl}. In this paper, the words ``system
potential'' will always refer explicitly to the renormalized system
potential. A physical restriction on the form of the couplings
$\Gamma_A$ arises from the requirement that the renormalized
potential not change the qualitative form of the bare
potential. Thus, e.g., if the bare potential is a polynomial of order
$n$, the renormalization should induce no terms of order higher
than $n$. This condition also insures stability of the system towards
the destabilizing effects of multiplicative noise, since it implies
that the stochastic forcing terms in the potential must be a
polynomial of order ${\le}{\;}n$.

Finally, as emphasized, e.g., by Caldeira and Leggett \cite{kn:cl},
it should be stressed that this is more than simply a toy model.
This form of the Hamiltonian generally provides a correct description
for {\em any} system which is only weakly coupled to its surroundings.
To facilitate a concrete calculation, this Hamiltonian need only be
supplemented by two inputs, namely the spectral distribution of the
environmental modes and the form of the coupling to the system. For a
general physical problem, these may either be extracted from
experimental data or derived from theoretical considerations.
\newpage

\centerline{\bf III. Time-Dependent Langevin Equations with
Multiplicative Noise}
\vskip .1in

The equations of motion generated from the Hamiltonian (\ref{6})
clearly take the forms
\ba
\dot{x}&=&v, \nonumber\\
\dot{v}&=&-{\pa\over\pa x}{\,}U(x,t)
+\sum_A {\Omega}_{A}^{~2}(t)\left[q_A-\Gamma_A(x,t)\right]
{\pa\over\pa x}\G_A(x,t), \nonumber\\
\dot{q}_A&=&p_A, \nonumber\\
\dot{p}_A&=&-{\Omega}_{A}^{~2}
\left[q_A-\Gamma_A(x,t)\right], \label{8a}
\ea
where an overdot denotes a time derivative ${\partial}/{\partial}t$.

The fact that the equation for $\dot{p}_A$ is linear in $q_A$
implies that one can immediately write down a formal solution for
$q_A(t)$ in terms of $\G_A(x,t)$ at retarded times $s<t$. Indeed, let
$S_A(t)$ and $C_A(t)$ denote two linearly independent solutions to
the homogeneous oscillator equation
\bq
\ddot{\Xi}_A+{\Omega}_{A}^{~2}{\Xi}_A(t)=0,  \label{9}
\eq
chosen without loss of generality to satisfy the initial conditions
$C_A(0)=\dot{S}_A(0)=1$ and $S_A(0)=\dot{C}_A(0)=0$ at some time
$t=0$. One then concludes exactly that
\ba
q_A(t)&=&q_A(0)C_A(t)+p_A(0)S_A(t)    \nonumber\\
&&+\int_0^{t}ds~{\Omega}_{A}^{~2}(s)\Gamma_A(x(s),s)
\left[S_A(t)C_A(s)-S_A(s)C_A(t)\right]. \label{10}
\ea
The integrand in (\ref{10}) vanishes in the coincidence
limit $s\rightarrow t$. This, however, may be remedied by replacing
$S_A$ and $C_A$ by $-\ddot{S}_A/{\Omega}_{A}^{~2}$ and
$-\ddot{C}_A/{\Omega}_{A}^{~2}$ and then integrating by parts. The net
result is a formal solution
\ba
q_A(t)-\Gamma_A(x(t),t)&=&\left[q_A(0)-\Gamma_A(x,0)\right]C_A(t)+
p_A(0)S_A(t)\nonumber\\
&&+\int_0^{t}ds~W_A(s,t){{\partial} \over {\partial}s}\Gamma_A(x(s),s),
\label{11}
\ea
where the Wronskian
\bq
W_A(s,t)\equiv \dot{S}_A(s)C_A(t)-\dot{C}_A(s)S_A(t).  \label{12}
\eq
Note that, for the special case of time-independent frequencies,
$C_{A}(t)=\cos{\Omega}_{A}t$ and
$S_{A}(t)={\Omega}_{A}^{-1}\cos{\Omega}_{A}t$.

By inserting (\ref{11}) into the equation for $\dot{v}$ and
grouping terms suggestively, one then recovers an exact, nonlocal
equation of the form
\bq
\dot{v}=-{{\partial}U\over {\partial}x} -
\int_0^{t}ds \sum_A {\Omega}_{A}^{~2}(t)W_A(s,t){{\partial} \over
{\partial}x}\Gamma_A(x(t),t)
{{\partial} \over {\partial}s}\Gamma_A(x(s),s) + F_s(t),  \label{13}
\eq
where
\bq
F_s(t)=\sum_A {\Omega}_{A}^{~2}(t) \left[{{\partial}\over {\partial}x}
\Gamma_A(x(t),t)\right]\left\{\left[q_A(0)-\Gamma_A(x(0),0)\right]C_A(t)
+ p_A(0)S_A(t)\right\}. \label{14}
\eq

In the spirit of the discussion in Section II, suppose now that the
interaction between system and bath entails a polynomial coupling
\bq
{\Gamma}_{A}(x,t)= \sum_{n=1}^{N} {1\over n}\gamma_{A}^{(n)}(t)x^{n},
\label{15}
\eq
where the functions $\gamma_{A}^{(n)}$ are arbitrary real functions of
time. Equation (\ref{13}) then takes the form
\bq
\dot{v}=-{{\partial}U\over {\partial}x}
-\int_0^{t}ds \left[K(t,s)v(s)+M(t,s)x(s)\right] + F_s(t), \label{16}
\eq
where, in terms of the quantities
\bq
A_A(s,t)=\left(\sum_m \gamma_{A}^{(m)}(s)x^{m-1}(s)\right)
         \left(\sum_n \gamma_{A}^{(n)}(t)x^{n-1}(t)\right) \label{17}
\eq
and
\bq
B_A(s,t)={1\over n}{{\partial}\over {\partial}s}A_A(t,s), \label{18}
\eq
the ``memory kernels'' $K(t,s)$ and $M(t,s)$ are
\ba
K(t,s)&=&\sum_A\Omega_A^{~2}(t)A_A(s,t)W_A(s,t), \label{19}\\
M(t,s)&=&\sum_A\Omega_A^{~2}(t)B_A(s,t)W_A(s,t).  \label{20}
\ea
The force $F_s$ now reduces to
\bq
F_s(t)=\sum_A {\Omega}_{A}^{~2}(t)
\sum_m \gamma_{A}^{(m)}(t)x^{m-1}
\left\{
\left[q_A(0)-\Gamma_{A}(x,0)\right]C_A(t)+p_A(0)S_A(t)\right\}.
\label{21}
\eq

Equation (\ref{16}) is considerably more complicated than an ordinary
Langevin equation. However, these additional complications need not
preclude entirely the possibility of a simple physical interpretation
or the proof of a fluctuation-dissipation theorem. Provided
that the oscillator frequencies ${\Omega}_{A}$ are not time-dependent,
one can still prove a fluctuation-dissipation theorem, even if
$\Gamma_A$ is a nonlinear function of $x$ \cite{kn:rz} and/or
explicitly time-dependent. Indeed, consider an ensemble of initial
conditions for which the first moments vanish identically, i.e.,
\bq
\VEV{Q_A(0)}=\VEV{{p_A}(0)}\equiv 0, \label{22}
\eq
with $Q_A(0)\equiv q_A(0)-\Gamma_A(x(0),0)$, and where the second
moments are initially thermal, so that
\bq
\VEV{{p_A(0)p_B(0)}}=
{\Omega}_{A}(0){\Omega}_{B}(0)\VEV{Q_A(0)Q_B(0)}
=k_BT\delta_{AB}, \label{23}
\eq
where the angular brackets denote an initial ensemble average. One
then computes exactly that
\bq
\VEV{F_s(t)}=0 \label{24a}
\eq
and
\bq
\VEV{F_s(t_1)F_s(t_2)}=k_BTK(t_1,t_2), \label{24}
\eq
thereby identifying $F_s(t)$ as a noise and proving a generalized
fluctuation-dissipation theorem linking the noise autocorrelator with
the ``viscosity kernel'' $K(t,s)$. (Strictly speaking $K(t,s)$ need give
rise to a true viscosity only in the case of an ohmic environment, a
point which will be discussed later.) Equation (\ref{16}) can now be
viewed as a nonlinear, nonlocal Langevin equation.

That a fluctuation-dissipation theorem can hold even in these more
complicated settings is a formal consequence of the fact that the
nonlinearities and time-dependences in (\ref{16}) enter into
the memory kernel $K(t_1,t_2)$ and the autocorrelator
$\VEV{F_s(t_1)F_s(t_2)}$ in exactly the same way. Physically, this
result can be understood as follows: If the basic picture is valid,
the total energy is dominated by the constant energy of the heat bath
(recall that one is now assuming that the bath frequencies are
time-independent), so that energy is approximately conserved, even
if the system Hamiltonian $H_S$ is time-dependent. Since fluctuations
induce a monotonic increase in the system energy, there must be some
source of dissipation if that energy is to remain bounded. However, if
that dissipation is too strong and dominates the fluctuations, the system
energy will vanish at late times, which is clearly unphysical for a
system coupled to a finite temperature heat bath. The fluctuations and
the dissipation must clearly balance if the system is to have a finite
but nonzero energy at late times.

It should be observed that the Langevin equation (\ref{16}) reduces to
a well known form in an appropriate limit. If one neglects all
nonlinearities in the coupling between system and bath, assuming that
$\Gamma_A\propto x$, one immediately recovers a special model
considered previously \cite{kn:hka}. And, moreover, if one assumes
further that ${\Omega}_{A}$ and $\Gamma_{A}$ are independent of time,
one is reduced to the well known independent oscillator model
\cite{kn:io}. It is thus possible to address systematically the
question of how the incorporation of nonlinearities and/or
time-dependences leads to systematic changes in the Langevin equation
derived for that original model.

When one neglects both the nonlinearities and the time dependences,
the memory kernel $M(s,t)$ vanishes identically and, moreover, the
remaining memory kernel $K(s,t)$ contains no explicit $x$-dependence.
The stochastic force $F_{s}$ involves $x$ only linearly, through the
propagation of an initial condition. It thus follows that one recovers a
relatively simple equation involving a (nonlocal) friction $\propto v(s)$
and purely additive noise. To the extent that the function $K(t,s)$ is
sufficiently sharply peaked about the coincidence limit $s \to t$ (what
precisely this entails will be discussed below), one can then approximate
$v(s)$ by its value at time $t$, in which case one obtains a Markovian
equation involving a friction $\propto v(t)$.

When one allows for a nontrivial time-dependence in the oscillator
frequencies or the coupling between the system and bath, but as yet no
nonlinearities, an additional nonvanishing kernel $M(s,t)$ appears.
In a Markov approximation, this leads to a new term in the
Langevin equation proportional to $x(t)$ which corresponds to a
time-dependent change in the system potential. In a field theoretic
context, this would be interpreted as a mass renormalization. In
this case, $F_{s}$ still gives rise to additive noise, but that noise
acquires an explicit time-dependence.

When instead one incorporates nonlinearities but no time-dependences,
the memory kernel $M(s,t)$ still vanishes, but the other kernel
$K(s,t)$ becomes significantly more complicated, involving not simply
an autocorrelator for the mode functions
$C_A\propto \cos{\Omega}_{A}t$, but a correlator of
${\partial}\Gamma_{A}/{\partial}x$ with itself. In other words, the
nonlinearity implies that the evolution of $x$ actually involves an
$x$-dependent memory. This is hardly surprising. Indeed, as one might
have anticipated, e.g., by analogy with the theory of the Brownian
motion, the integral is nothing other than the autocorrelation
function for the forces associated with the interaction of the system
with each of the bath modes. In this case, $F_{s}$ also acquires an
additional $x$-dependence implying that the noise will depend not only
on the initial conditions $x(0)$, $q_A(0)$, and $p_A(0)$, but upon the
$x$-dependent state of the system as well. In other words, the noise
is multiplicative.

It should be observed that the memory kernel $M(t,s)$ does not enter
into the fluctuation-dissipation theorem for a time-independent bath.
Its only effect is to induce a systematic renormalization of the
system potential, which will of course affect the {\em form} of the
late time solution.

One further point should be stressed. The fluctuation-dissipation
theorem of equation (\ref{24}) refers explicitly only to the total force
$F_{s}$ and the total memory kernel $K(t,s)$. However, it is easy to
see that analogous theorems also hold separately for each term
$\propto \g_{A}^{(m)}(s)\g_{A}^{(n)}(t)$. In this sense, the
fluctuation-dissipation theorem is truly microscopic.

Equation (\ref{16}) is an exact, nonlocal equation. Only
to the extent that this equation can be approximated as Markovian
can one derive from it a Fokker-Planck equation. Such a Markov
limit implies (1) that the memory kernels $K(t,s)$ and $M(t,s)$ may be
approximated as essentially local in time, and (2) that the
autocorrelation function for $F_s$ falls off rapidly as $|t-s|$
increases. It is clear by inspection that, if $K(t,s)$ is essentially
local, so is $M(t,s)$. Further, the fluctuation-dissipation theorem
guarantees that, if $K(t,s)$ is local, the noise autocorrelation will
be as well. In considering the validity of a Markov approximation, it
thus suffices to consider $K(t,s)$.

The Markov limit can oftentimes be justified approximately
for smooth spectral distributions, given an appropriate separation
of time scales. Let $t_{C}$ denote a characteristic time scale on
which the memory kernels $K(t,s)$ and $M(t,s)$ decay, and let $t_{sys}$
denote a characteristic time scale on which the system velocity $v$
and/or position $x$ change significantly, either in response to the
environment ($t_{R}$) or to the system potential $U(t_{S})$.
(Provided that the mode functions are all oscillatory, $K(t,s)$ will
decay much faster than $|t-s|$ at late times, so that one can clearly
identify a time scale $t_{C}.$) In a time-independent setting, the
Markov limit then follows when $t_{C}{\;}{\ll}{\;}t_{sys}$. If the
oscillators are time-dependent, there is another relevant time scale,
$t_{H}$, the characteristic time on which the frequencies change. To
the extent that $t_{H}$ is much larger than both $t_{sys}$ and
$t_{C}$, one anticipates that the time-dependence may be viewed as a
perturbation, and that the Markov limit should still obtain. If
$t_{C}{\;}{\ll}{\;}t_{sys}$, but is {\em not} short compared with
$t_{H}$, the time dependence can no longer be viewed as a
perturbation, but it may still be true that a Markov approximation can
be justified. This should, e.g., be the case if all the modes still
oscillate and/or the coupling of the system to the longest wavelength
modes is relatively weak. If, however, $t_{H}$ is not much larger than
$t_{C}$, one expects that a local Fokker-Planck description will be
inappropriate. This is, for example, true for the specific example
discussed in Sections IV and V.

For the special case of ``ohmic'' environments, the nonlocal term in
the exact Langevin equation involving $K(t,s)$ reduces to the usual
linear viscosity seen in heuristic {\em ad hoc} Langevin descriptions.
These environments are characterized by time-independent frequencies and
have a spectral distribution $g(\Omega) \propto \Omega^{2}$,
with an upper cut off at some ${\Omega}_{max}$, and all the oscillators are
assumed to couple bilinearly to the system with an equal strength
\cite{kn:cl} \cite{kn:rz}. If the coupling is not bilinear, one still
recovers a viscosity that is linear in $v$, but this viscosity will
dependent explicitly on $x$. If the spectral distribution
$g({\Omega})$ differs only slightly from $\propto \Omega^{2}$, that
difference may be treated perturbatively to extract calculable
modifications in the form of the Markovian equation. However, for
spectral distributions which are very different, e.g., ``supra-ohmic''
distributions $\propto \Omega^{4}$, the local Langevin equation
can be higher order in time derivatives and need not contain a simple
viscosity term $\propto v(t)$. One concrete example thereof is
provided by an electron interacting with its self-electromagnetic
field \cite{kn:ac}. However, for the cosmological example considered
in Sections IV and V, the spectral distribution will be nearly
``ohmic,'' so that the local Langevin description will contain an
ordinary viscosity and thus admit a Fokker-Planck realization.
\newpage

\centerline{\bf IV. A Cosmological Example}
\vskip .1in
The objective of this Section is to formulate a nonlocal Langevin
equation in terms of which to describe the evolution of some system
variable, evolving in a Landau-Ginsburg potential $V_{ren}$ and
coupled to a bath of scalar black body ``radiation.'' The aim of this
Langevin equation is to provide a quasi-realistic model for a
cosmological phase transition. The entire analysis will be classical,
allowing for thermal fluctuations but {\em not} for quantum
fluctuations. This may prove appropriate either in the context of some
versions of inflation or, alternatively, in the formation of
cosmic strings or baryogenesis. Attention here focuses on
formulating the problem and discussing its physical potentialities.
Quantitative details will be provided in a subsequent paper.

In what follows the bath will be idealized as a collection of
oscillators, characterized by time-dependent frequencies appropriate
for a scalar field with a general ${\xi}R$ curvature coupling. The
case  ${\xi}=1/6$ corresponds to conformal coupling, whereas ${\xi}=0$
yields minimal coupling. Consistent with the discussion in Section II,
the interaction Hamiltonian will be taken as linear in the oscillator
variables $q_{A}$, but it can involve an arbitrary quadratic
dependence on the system variable $x$, with both linear and nonlinear
pieces. By allowing for both linear and nonlinear
couplings, and varying the relative strengths of these two different
contributions, one will be able to compare the effects of additive and
multiplicative noise. The analysis will be effected in the conformal
frame, in terms of a conformal time coordinate ${\eta}$ satisfying
$d{\eta}=a^{-1}dt$, where $a$ denotes the scale factor. It will,
moreover, be assumed that the spatial curvature of the $t=$ constant
slices vanishes, so that one is considering a $k=0$ Friedmann
cosmology.

Given these assumptions, one is led directly to a Hamiltonian of the
form motivated in Section II, namely
\bq
H={1\over 2}v^{2}+U(x,{\eta}) +
{1\over 2}\sum_{A}\left\{ p_{A}^{~2}+
{\Omega}_{A}^{~2}({\eta})\left[q_{A}-\Gamma_{A}(x,
{\eta})\right]\right\}^{2}, \label{25}
\eq
where, in terms of constants $\lambda$, $\theta$, and $\sigma$,
\bq
U(x,{\eta})={1\over 4}\l ({\eta})x^{4}+
{1\over 3}{\theta}({\eta}) x^{3} +
{1\over 2}{\sigma}({\eta}) x^{2}, \label{26}
\eq
and the coupling
\bq
\Gamma_{A}(x,{\eta})={\gamma}^{(1)}_{A}({\eta})x +
{\gamma}^{(2)}_{A}({\eta})x^{2}. \label{27}
\eq
Recall from equation (\ref{7}) that $U$ and the Landau-Ginsburg
potential $V_{ren}$ are connected by a term involving the couplings
${\Gamma}_{A}$. It follows that, if the ${\gamma}_{A}^{(1)}$'s and
${\gamma_{A}}^{(2)}$'s are both nonvanishing, this coupling will in
general induce a cubic term in $U$, although the coefficient $\theta$
of that term {\em could} vanish. The coefficients ${\lambda}$ and
${\sigma}$ in $U$ are also ``dressed'' quantities, differing from the
bare quantities in $V_{ren}$ because of the terms involving the
${\Gamma}_{A}$'s. At this stage, the explicit time dependence of
${\lambda}$ and ${\sigma}$ and the couplings ${\gamma}^{(1)}_{A}$ and
${\gamma}^{(2)}_{A}$ may be treated  as more or less arbitrary.
The frequencies ${\Omega}_{A}$ satisfy
\bq
{\Omega}_{A}^{~2}({\eta})={\omega}_{A}^{~2}-(1-6{\xi}){a''\over a}
\label{28}
\eq
where a prime $'$ denotes a conformal time derivative
${\partial}/{\partial}{\eta}$.

Suppose now that, in terms of cosmic time $t$, the scale factor $a$
evidences a simple power law time dependence $a=t^{p}$, with $p\geq
1/2$. It then follows that
\bq
a{\,}{\propto}{\,}({\eta}-{\eta}_{0})^{p/(1-p)}, \label{29}
\eq
where ${\eta}_{0}$ denotes an integration constant, so that
\bq
{\Omega}_{A}^{~2}({\eta})={\omega}_{A}^{~2}-
{\nu^{2}\over ({\eta}-{\eta}_{0})^{2}}, \label{30}
\eq
where
\bq
\nu^{2}=(1-6{\xi}){p(2p-1)\over (1-p)^{2}} \label{31}
\eq
is intrinsically positive when $p>1/2$ and $1-6\xi \geq 0$.
The quantities ${\omega}_{A}^{~2}$ denote eigenvalues of the spatial
Laplacian.

To proceed further, one needs to determine the mode functions
${\Xi}_{A}$ for the time-dependent frequencies. These can clearly be
evaluated in terms of Bessel functions. However, in so doing there
are at least two possible ways in which to proceed, namely considering
(1) complex modes involving H\"ankel functions or (2) real modes
involving ordinary Bessel and Neumann functions. This paper will adopt
the (less conventional) second choice, since it provides
for a more direct connection with earlier work on Langevin equations:
in the absence of a time-dependent expansion, the mode functions
reduce to sines and cosines. Straightforward algebra reveals that
the equation
\bq
{\Xi}_{A}'' +\left[{\omega}_{A}^{~2} - {\nu^{2}\over
\left({\eta}-{\eta}_{0}\right)^{2}}\right]{\Xi}_{A} = 0 \label{32}
\eq
is solved by a general
\bq
{\Xi}_{A}=\left[{\omega}_{A}\left({\eta}-
{\eta}_{0}\right)\right]^{1/2}
Z_{\mu}\left({\omega}_{A}\left({\eta}-{\eta}_{0}\right)\right),
\label{33}
\eq
where $Z_{\mu}$ denotes an arbitrary solution to Bessel's equation of
order
\bq
{\mu}^{2}=(1-6{\xi}){\,}{p(2p-1)\over (1-p)^{2}} + {1\over 4}=
{\nu}^{2}+ {1\over 4}, \label{34}
\eq

Note that, for the special cases ${\xi}=1/6$ (conformal coupling)
and/or  $p=1/2$ (a Universe dominated by conformal electromagnetic
radiation), ${\mu}=1/2$ and the solutions ${\Xi}_{A}$ reduce to sines
and cosines. For these particular values, the bath Hamiltonian
${\delta}H_{B}$ is time-independent in the conformal frame. This
implies that the physical frequencies are simply red-shifted uniformly
as the Universe expands.

Presuming that the initial value problem for the Langevin equation is
specified at time ${\eta}=0$, the appropriate solutions will be
$C_{A}$ and $S_{A}$, combinations of Bessel and Neumann functions
$J_{\nu}$ and $N_{\nu}$, modulated by factors
${\omega}_{A}\left({\eta}-{\eta}_{0}\right)$, satisfying
$C_{A}(0)=dS_{A}(0)/d{\eta}=1$ and $S_{A}(0)=dC_{A}(0)/d{\eta}=0$.
If ${\mu}{\ne}1/2$, the functions $C_{A}$ and $S_{A}$ are, for long
wavelengths with $\abs{\omega_A\left({\eta}-{\eta}_{0}\right)} \ll \mu$,
very different from sines and cosines. Indeed, they are not even
oscillatory. However, to the extent that such longer wavelengths do
not couple significantly to the system, one can approximate the
${\Xi}_{A}$'s by the forms appropriate when
$\abs{\omega_A\left({\eta}-{\eta}_{0}\right)} \gg \mu$. In this case,
the mode functions reduce to
\bq
{\Xi}_{A}({\eta}) \approx \left({2\over \pi}\right)^{1/2}
\sin\left[\omega_A\left({\eta}-{\eta}_{0}\right)-{\pi \over
2}\left({\mu}+{1\over 2}\right)\right] \label{35}
\eq
\bq
{\Xi}_{A}({\eta}) \approx \left({2\over \pi}\right)^{1/2}
\cos\left[\omega_A\left({\eta}-{\eta}_{0}\right)-{\pi \over
2}\left({\mu}+{1\over 2}\right)\right], \label{36}
\eq
i.e., ordinary sines and cosines, modulated by phase shifts which can
of course be absorbed in the normalizations.

It is clear that, in this limit, one recovers a relatively simple
nonlocal Langevin equation, for which the only explicit time
dependences are in the potential $U$ and the couplings
between the system and the environment. It thus follows that, in
this approximation, a fluctuation-dissipation theorem holds, so that
one would anticipate an evolution towards some steady state solution
at late times. The fluctuation-dissipation theorem is of course exact
when $\xi=1/6$ and/or $p=1/2$.

It should be stressed that the condition
$\abs{\omega_A({\eta}-{\eta}_{0})} \gg \mu$ has a very simple physical
interpretation. Reexpressed in terms of the physical cosmic time $t$,
this condition becomes $({\omega}_{A}/a)t \gg \mu\abs{1-p}$, this
corresponding, for $\mu$ and $p$ of order unity, to the demand that the
physical period of the oscillation be short compared with the time
scale $t_{H}$ associated with the expansion of the Universe. In other
words, the wavelength must be short compared with the horizon length.

In general, however, a fluctuation-dissipation theorem does not hold.
Recall that the memory kernel $K(t,s)$ satisfies
\bq
K(t,s)=\sum_A\Omega_A^{~2}(t)A_A(s,t)W_A(s,t), \label{37}
\eq
in terms of the Wronskian of equation (\ref{12}). Alternatively, a
thermal average of the noise autocorrelation, taken at the initial
time ${\eta}=0$, satisfies
\bq
\left(k_{B}T\right)^{-1}\VEV{F_{s}(t)F_{s}(s)}=
\sum_A {\Omega}_{A}^{~2}(t)A_A(s,t)Q_A(s,t), \label{38}
\eq
where
\bq
Q_A(s,t)={\Omega}_{A}^{~2}(s)\left[
{C_{A}(t)C_{A}(s)\over {\Omega}_{A}^{~2}(0)} +
S_{A}(t)S_{A}(s)\right]. \label{39}
\eq
When the frequencies ${\Omega}_{A}$ are all time-independent
($\xi=1/6$ or $p=1/2$), $W_{A}(s,t)$ and $Q_{A}(s,t)$ are in fact
equal, so that one recovers the fluctuation-dissipation theorem
(\ref{24}). In general, however, this is clearly not so; and the
typical size of the fractional deviation between these two quantities
provides a concrete measure of the degree to which the
fluctuation-dissipation theorem fails. Suppose that ${\omega}_{cr}$
denotes a typical frequency associated with the coupling of the system
and the environment. It then follows straightforwardly that the
fractional amplitude of the deviation is of order
\bq
{1\over {\omega}_{cr}\left({\eta}-{\eta}_{0}\right)}~{\sim}~
{a\over {\omega}_{cr}t}. \label{40}
\eq
The fractional deviation from a fluctuation-dissipation theorem scales
as the ratio of a characteristic oscillator period
$\sim ({\omega}_{cr}/a)^{-1}$ to the expansion time scale $t_{H}$.

It remains to consider the circumstances under which a Markov
approximation can be justified. Note first of all that, to the extent
that the bath oscillators are interpreted as representing the modes of
some free field, one would anticipate a spectral distribution
$\propto \omega_A^{~2}$, so that, given some cutoff
${\omega}_{max}$, one can pass to a continuum limit
\bq
\sum_{A} \to \int d\omega~\omega^2. \label{41}
\eq
Consistent with the equation for ${\delta}H_{I}$, now separate out the
explicit ${\Omega}$-dependence in the system-bath couplings and write
${\gamma}_{A}^{(n)}(\eta)=c_{A}^{(n)}(\eta)/{\Omega}_{A}^{~2}({\eta})$.

As a particularly simple first approximation, suppose that
$c^{(n)}_{A}$ is essentially independent of frequency.
And, moreover, assume that the longest wavelength modes are not very
important in the coupling, so that one can neglect the frequency shift
associated with the expansion of the Universe and set
$\Omega_A \approx \omega_A$. Given these approximations, one has,
e.g., that
\bq
K(\eta,s){\;}{\approx}{\;}\int d{\omega} \sum_{m,n}
c^{(n)}(s)c^{(m)}(\eta) x^{n-1}(s)x^{m-1}(\eta)
\cos{\omega}(\eta -s). \label{42}
\eq
Suppose, however, that one can also neglect the time-dependence of the
$c$'s, a reasonable assumption, e.g., {\em if}\/ the $c$'s change only
on an expansion time scale $t_{H}$. One can then effect the
$d{\omega}$ integration explicitly to obtain (for large
${\omega}_{max}$) a delta function ${\delta}_{D}(\eta-s)$.
It follows that the memory of $K(t,s)$ is indeed very short, so that,
presuming that the time scale $t_{S}$ associated with $U$ is not too
short, one can approximate
\bq
K(\eta,s){\;}{\approx}{\;}\left[\sum_n c^{(n)}(t)x^{n-1}(\eta)
\right]^{2}{\delta}_{D}(\eta -s){\;}{\equiv}{\;}
2{\cal K}(x,\eta){\delta}_{D}(\eta -s). \label{43}
\eq
An analogous expression holds for the other memory kernel $M(\eta,s)$.

If the $c_{A}$'s depend strongly on frequency and/or the spectral
distribution is considerably different, the analysis becomes more
complicated. It is, however, evident that, whenever the couplings are
such that the lowest frequency modes are not very important, so that
$t_{C} \ll t_{R}$ and $t_{H}$, and the distribution of modes is
approximately ohmic, a Markov approximation should in fact be
justified, leading to a viscosity ${\propto}{\;}{\cal K}v$.

Under these circumstances, the nonlocal Langevin equation can be well
approximated by a local equation of the form
\bq
v'=-{{\partial}U\over {\partial}x} -
{\cal M}(\eta)x(\eta) - {\cal K}(\eta)v(\eta) + F_{s}(\eta), \label{44}
\eq
where, for an initial thermal ensemble,
\bq
{\langle}F_{s}(\eta)F_{s}(s){\rangle}=
2k_{B}T{\cal K}(x,{\eta}){\delta}_{D}(\eta-s) \label{45}
\eq
This local Langevin equation leads immediately to a Fokker-Planck
equation of the form
\bq
{{\partial}f\over {\partial}{\eta}} +
{{\partial}\over {\partial}x}(vf) +
{{\partial}\over {\partial}v}\left[\left(-{\pa U\over\pa x} - {\cal
M}x - {\cal K}v\right)f\right] - k_{B}T{\cal
K}{\,}{{\partial}^{2}f\over {\partial}v^{2}}=0. \label{46}
\eq

Translated back into the physical frame, the Langevin equation
(\ref{44}) becomes
\bq
\dot{V}=-{1\over a^4}{\pa U\over\pa X}-{\bf
M}(t)X-\left[3{\dot{a}\over a}+{\bf K}(t)\right]V+{\bf F}_s(t), \label{47}
\eq
where the overdot denotes differentiation with respect to cosmic
time, and
\ba
X&=&{x\over a},               \label{a1}\\
V&=&\dot{X}={v\over a^2}-{\dot{a}\over a^2}x, \label{a2}\\
{\bf K}(t)&=&{{\cal K}\over a},               \label{a3}\\
{\bf M}(t)&=&{{\cal M}\over a^2}+{\bf K}(t)\left({\dot{a}\over
a}\right)+
\left({\dot{a}\over a}\right)^2+{\ddot{a}\over a},
\label{a4}\\
{\bf F}_s(t)&=&{F_s\over a^3}.  \label{a5}
\ea
The noise autocorrelator is now
\bq
\VEV{{\bf F}_s(t){\bf F}_s(t')}=2{k_BT\over a^4}{\,}{\bf K}(t)\d(t-t').
\label{a6}
\eq
Note that, in the physical frame, there are two sources of
damping, namely the viscosity $\propto {\bf K}V$ and the
cosmological frame-dragging $\propto H\equiv\dot{a}/a$.
The corresponding Fokker-Planck equation is now
\bq
{\pa f_p\over\pa t}+{\pa\over\pa X}(Vf_p)+{\pa\over\pa
V}\left\{\left[-{1\over a^4}{\pa U\over\pa X}-{\bf
M}(t)X-\left(3{\dot{a}\over a}+{\bf
K}(t)\right)V\right]f_p\right\}-{k_BT\over a^4}{\bf
K}(t){\pa^2f_p\over\pa V^2}=0.      \label{a7}
\eq

The form of this equation is consistent with the Fokker-Planck equation
analyzed by Brandenberger {\em et al} \cite{kn:rba}. Given, however, that
the approach followed here is entirely different from that of Ref.
\cite{kn:rba} a more detailed comparison is appropriate:
The present formalism is intended primarily to describe the behavior
of a homogeneous degree of freedom (dilaton mode). However, for the special
case of a free field with no mode couplings, the Fokker-Planck equation
(\ref{a7}) holds equally well for {\em any} single field mode coupled to
a heat bath, which is precisely the model problem considered by
Brandenberger {\em et al}. Comparing (\ref{a7}) with equation (20) of
Ref. \cite{kn:rba}, one finds that, even though the latter equation
was arrived upon in a rather different way, the diffusion terms are
consistent in that they scale the same way with $a$, and that in both
cases the ``Hubble damping'' term is present. However there {\em is}
one important difference. While the present model incorporates
back-reaction effects due to the heat bath, namely the normal
viscosity ${\bf K}$ and the potential renormalization ${\bf M}$, these
physical effects are not taken into account in Ref. \cite{kn:rba}. A
quantitative assessment of the importance of these effects will be
presented elsewhere.

Having formulated the Fokker-Planck equation, it is worth recalling
once again the critical assumptions that went into its derivation. (1)
The decay time $t_{C}$ must be short compared with the time scales
$t_{R}$ and $t_{S}$ on which the system changes in response either to
its potential $U$ or in response to the environment {\em and} with the
expansion time $t_{H}$. (2) For
${\omega}({\eta}-{\eta}_{0}){\;}{\ll}{\;}{\mu}$, the modes are
non-oscillatory. One must also assume that the coupling of the system
with these infrared modes is negligible, so that there is no
significant long time tail to $K({\eta},s)$ to prevent the existence
of a Markov limit. (3) The spectral distribution must be approximately
``ohmic,'' with $g({\omega}) \propto \omega^{2}$, so that the nonlocal
contribution involving $K({\eta},s)v(s)$ gives rise to an ordinary
viscosity. Fortunately, this is precisely what one expects of a heat
bath of thermal photons.
\newpage

\centerline{\bf V. An Approximate Solution}
\vskip .1in
Given the assumed form (\ref{25}) for the Hamiltonian (with the
system-bath coupling assumed to be weakly time-dependent), the
coefficients ${\cal K}$ and ${\cal M}$ may be written as
\bq
{\cal K}={\lambda}_{0}+2{\lambda}_{1}x+{\lambda}_{2}x^{2}
\eq
and
\bq
{\cal M}={\mu}_{0}+2{\mu}_{1}x+{\mu}_{2}x^{2}  \label{48}
\eq
and the noise correlator
\bq
\VEV{F_{s}({\eta})F_{s}(s)}
=2k_{B}T{\cal K}{\delta}_{D}({\eta}-s)
=2k_{B}T\left[
{\lambda}_{0}+2{\lambda}_{1}x+{\lambda}_{2}x^{2}\right]
{\delta}_{D}({\eta}-s),  \label{49}
\eq
where the ${\lambda}_{n}$'s and ${\mu}_{n}$'s are independent of $x$
and $v$. Note that ${\lambda}_{0}$ and ${\lambda}_{2}$ are necessarily
positive definite, but that ${\lambda}_{1}$ and the ${\mu}_{n}$'s are of
indeterminate sign. ${\cal K}$ gives rise to an $x$-dependent viscosity,
whereas ${\cal M}$ induces a further renormalization of the potential $U$
in terms of new coefficients ${\Lambda}$, ${\Theta}$, and ${\Sigma}$:
\ba
{\cal U}(x,{\eta})&=&
{1\over 4}{\lambda}({\eta}) x^{4}+
{1\over 3}{\theta}({\eta}) x^{3} +
{1\over 2}{\sigma}({\eta}) x^{2} +
{1\over 2}{\mu}_{0}x^{2}+{2\over 3}{\mu}_{1}x^{3}+{1\over
4}{\mu}_{2}x^{4} \nonumber\\
&\equiv&{1\over 4}{\Lambda}({\eta}) x^{4}+
{1\over 3}{\Theta}({\eta}) x^{3} +
{1\over 2}{\Sigma}({\eta}) x^{2} \label{50}
\ea

By inserting (\ref{48}) and (\ref{49}) into (\ref{46}), one obtains an
explicit Fokker-Planck equation, which one may hope to solve.
Unfortunately, however, in general such a multivariable Fokker-Planck
equation cannot be solved exactly except via numerical techniques.
Nevertheless, one {\em can} at least obtain an approximate solution
for the expectation value $\VEV{E}$ of the system energy $E$ using the
so-called ``energy envelope'' technique introduced by Stratonovich
\cite{kn:strat} and further developed by Lindenberg and
Seshadri \cite{kn:ls}.

The idea underlying this approximation is in fact straightforward: In
many cases of physical interest, such as that considered here, one can
visualize the system as exhibiting (nonlinear) oscillations on the time
scale $t_{S}~{\sim}~{\omega}_{0}^{-1}$ associated with the potential
${\cal U}$, oscillations which are eventually altered by the coupling
with the environment on the damping time scale $t_{R}$. To the extent
that the damping time $t_{R}$ and the expansion time $t_{H}$
are both long compared with $t_{S}$, one can then assume
that the system energy $E$ is nearly conserved on time scales
$\sim t_{S}$, and varies only on a time scale much longer than the
time associated with variations in $x$. It is thus natural to
transform from $x$ and $v$ to new variables $x$ and $E$, to treat $E$
as an adiabatic invariant, and to implement an ``orbit average'' of
the transformed $E$-$x$ Fokker-Planck equation to extract an equation
involving only $E$ and ${\eta}$.

Implement, therefore, a change of variables from $(x,v)$ to $(x,E)$,
where, explicitly. $E={1\over 2}v^{2}+{\cal U}(x,{\eta})$, to obtain a
new Fokker-Planck equation for $W(x,E,t)$ satisfying
\bq
f(x,v,\eta)dxdv=W(x,E,\eta)dxdE. \label{51}
\eq
To the extent that the energy is approximately conserved during a
single oscillation of the system, one can then assume that
\bq
W(x,E,\eta)=\{2{\Phi}'(E)\left[E-{\cal U}(x,\eta)\right]^{1/2}\}^{-1},
W_{1}(E,\eta)
\label{52}
\eq
where
\bq
{\Phi}(E)=\int dx\left[E-{\cal U}(x,{\eta})\right]^{1/2} \label{53}
\eq
and $'$ now denotes a ${\partial}/{\partial}E$ derivative. Here the
integration extends over the values of $x$ along the unperturbed
orbit associated with $E$. Note that the prefactor of $W_{1}$ is simply
the relative amount of time that, for fixed $E$, the system spends at each
point $x$.

By integrating the Fokker-Planck equation for $W(x,E,\eta)$ over $x$, one
obtains the desired equation for $W_{1}(E,{\eta})$, which takes the
form
\ba
&&
{\pa\over\pa \eta}W_{1}(E,{\eta})=\nonumber\\
&&-\left({\pa\over\pa
E}\left[{1\over\chi'(E)}\left\{{\lambda}_{0}\left({\Phi}(E)-
k_BT{\Phi}'(E)\right) +
2{\lambda}_{1}\left({\chi}(E)-k_BT{\chi}'(E)\right)\right.\right.\right.
\nonumber\\
&&\left.\left.\left.+{\lambda}_{2}\left({\Psi}(E)-
k_BT{\Psi}'(E)\right)\right\}\right]\right.\nonumber\\
&&\left. +k_{B}T{{\partial}^{2}\over
{\partial}E^{2}}\left[{1\over\chi'(E)}\left\{{\lambda}_{0}{\Phi}(E) +
2{\lambda}_{1}{\chi}(E) +{\lambda}_{2}{\Psi}(E)\right\}\right]\right)
W_{1}(E,{\eta}), \label{54}
\ea
where
\bq
{\chi}(E)=\int {\,}dx{\,}x{\,}\left[E-{\cal U}(x,{\eta})\right]^{1/2}
\label{55}
\eq
and
\bq
{\Psi}(E)=\int {\,}dx{\,}x^{2}{\,}\left[E-{\cal U}(x,{\eta})\right]^{1/2}.
\label{56}
\eq
Note that this differs from equation (3.8) in \cite{kn:ls}, which assumes
implicitly that the system's unperturbed orbit is symmetric about
$x=0$, so that (\ref{55}) vanishes identically.

Unfortunately, for a generic potential ${\cal U}$ the functions
${\Phi}(E)$, ${\chi}(E)$, and ${\Psi}(E)$ cannot be evaluated
analytically, so that one cannot realize the right hand side of
(\ref{54}) explicitly in terms of simple functions of $E$. Thus, e.g.,
for the quartic potential (\ref{51}), these functions can only be
expressed as elliptic integrals, which must be evaluated numerically.
There {\em is}, however, one limit in which one can proceed
analytically, namely when the energy $E$ is small and the system is
oscillating about a local minimum of ${\cal U}$. The obvious point is
that, in this limit, one can evaluate the orbit integrals, assuming
that the system is effectively evolving in a simple harmonic
oscillator potential.

Suppose for simplicity that ${\Theta}=0$, so that the dressed
potential ${\cal U}$ is itself of the Landau-Ginsburg form. When
${\Sigma}$ is positive, one can then approximate the system as
oscillating with squared frequency $\Omega_0^{~2}=\Sigma$ about
the origin. And similarly, when ${\Sigma}$ is negative, and ${\cal U}$
is a ``Mexican hat'' potential, the system can be assumed to oscillate
with squared frequency $\Omega_0^{~2}=2\abs{\Sigma}$ about one of the
two minima at $x_{0}={\pm}({\Lambda}/\abs{\Sigma})^{1/2}$.

In either case, one can evaluate ${\Phi}$, ${\chi}$, and $\Psi$, to
realize the right hand side of (\ref{54}) in terms of polynomials at
most quadratic in $E$. And, given such an explicit representation, it
is straightforward to derive from the Fokker-Planck equation a
transport equation involving the time derivative of the first energy
moment
\bq
\VEV{E({\eta})}{\equiv}\int dE~E~W_{1}(E,{\eta}). \label{57}
\eq

Suppose in the first instance that ${\Sigma}$ is positive, and that
the system is executing small oscillations about the ground state
$x=0$ with ${\omega}_{0}^{~2}={\Sigma}$. In this approximation, one
verifies that, to the extent that the time dependence of $E$,
${\omega}_{0}$, and the ${\lambda}_{n}$'s may be ignored, the moment
equation takes the form
\bq
{\pa\over\pa\eta}\VEV{E({\eta})} =
k_{B}T{\lambda}_{0}-
\left({\lambda}_{0}-{k_{B}T{\lambda}_{2}\over
{\omega}_{0}^{~2}}\right)\VEV{E({\eta})}-
{{\lambda}_{2}\over {\omega}_{0}^{~2}}
\VEV{E^{2}({\eta})}. \label{58}
\eq
Note that, because of the reflection symmetry $x \to -x$ for the
potential ${\cal U}$, the functions ${\chi}={\chi}'=0$, so that the
contributions involving ${\lambda}_{1}$ vanish identically.

Unfortunately, this equation still cannot be solved exactly for
$\VEV{E({\eta})}$, as it involves the unknown function
$\VEV{E^{2}({\eta})}$. To obtain a formula for $\VEV{E^{2}({\eta})}$,
one must consider the second moment equation, which in turn relates
$d\VEV{E^{2}({\eta})}/d{\eta}$ to the third moment
$\VEV{E^{3}({\eta})}$. In the spirit of (say) the BBGKY hierarchy, one
requires a truncation approximation.

As in Refs. \cite{kn:strat} and \cite{kn:ls}, suppose therefore that
\bq
\VEV{E^{2}(t)}~\approx~\kappa\VEV{E(t)}^{2}, \label{59}
\eq
with ${\kappa}=2$. One knows that, when the system is at equilibrium,
with energy $E=k_{B}T$, this equation is satisfied identically for
${\kappa}=2$, and one might expect on physical grounds that, before the
system is ``at equilibrium,'' the energy distribution will be narrower
and ${\kappa}<2$. As emphasized by Lindenberg and Seshadri
\cite{kn:ls}, this truncation approximation thus leads to an upper
limit on the time scale on which the system ``equilibrates'' with the
bath. Given this truncation, one can immediately write down the
solution \cite{kn:ls}
\bq
\VEV{E({\eta})}=
{k_{B}T\left(E_{0}+Ak_{B}T\right) -
AkT\left(k_{B}T-E_{0}\right)\exp\left\{
-\left[(A+1)/A\right]{\lambda}_{0}{\eta} \right\}
\over \left(E_{0}+Ak_{B}T\right)-\left(k_{B}T-E_{0}\right)\exp\left\{
-\left[(A+1)/A\right]{\lambda}_{0}{\eta} \right\} } ,
\label{60}
\eq
where
\bq
A \equiv {\l_0\omega_0^{~2}\over{\lambda}_{2}k_{B}T} .
\label{61}
\eq

In the limit that ${\lambda}_{2} \to 0$, the multiplicative noise
``turns off'' and the system approaches an ``equilibrium'' with
$\VEV{E}=k_{B}T$ on a time scale $t_{R} \sim \lambda_0^{-1}$. If
${\lambda}_{2} \neq 0$, the system still evolves towards an equilibrium
with $\VEV{E}=k_{B}T$, but the time scale $t_{R}$ can be altered
significantly. Indeed, in the limit that ${\lambda}_{0} \to 0$, the
additive noise ``turns off'' and $t_{R} \sim
\omega_0^{~2}/({\lambda}_{2}k_{B}T)$. It thus follows that, when the
nonlinear coupling is sufficiently strong, the system may be driven
towards equilibrium, {\em not} by the ordinary additive noise
associated with the linear coupling, but primarily by the
multiplicative noise associated with the nonlinear coupling.

The only point that remains to be checked is that one is
still assuming, as is implicit in this ``envelope'' approximation,
that the time scale ${\omega}_{0}^{-1}$ is much shorter than the
damping time. This, however, is clearly the case when ${\lambda}_{0}$
and ${\lambda}_{2}$ are not too large.  Indeed, one verifies that (a)
the weak damping approximation {\em is} legitimate but (b)
multiplicative noise dominates the evolution towards an equilibrium
whenever \cite{kn:ls}
\bq
{{\lambda}_{0}\over {\omega}_{0}}{\;}{\ll}{\;}
{{\lambda}_{0}{\omega}_{0}^{~2}\over {\lambda}_{2}k_{B}T}
{\;}{\ll}{\;} 1 .  \label{62}
\eq

Turn now to the case when ${\Sigma}<0$ and the system is
oscillating about one of the two minima of the potential
$x_{0}{\ne}0$. This is the case relevant to first order phase
transitions. Here the terms involving ${\chi}$ and its
energy  derivative do not vanish, and the formulae for ${\Phi}$ and
${\Psi}$ acquire additional terms involving the location $x_{0}$ of
the new minimum. However, one still recovers a relatively simple exact
equation for ${\partial}\VEV{E}/{\partial}{\eta}$. Specifically, one
finds that, in this case, (\ref{58}) is replaced by
\bq
{\pa\over\pa\eta}\VEV{E({\eta})}=
k_{B}TL-\left(L-{k_{B}T{\lambda}_{2}\over
{\omega}_{0}^{~2}}\right)\VEV{{\cal E}({\eta})}-
{{\lambda}_{2}\over 2{\omega}_{0}^{~2}}
\VEV{{\cal E}^{2}({\eta})}, \label{63}
\eq
where now
\bq
{\cal E}=E-{\cal U}(x_{0}) \label{64}
\eq
denotes the system energy defined relative to the minimum of the
potential, and
\bq
L={\lambda}_{0}+{\lambda}_{1}x_{0}+{\lambda}_{0}x_{0}^{~2} \label{65}
\eq
plays the role of a ``dressed'' ${\lambda}_{0}$. The obvious point
here is that, since one is effectively expanding in a Taylor series
around the point $x_{0}={\pm}({\Lambda}/\abs{\Sigma})^{1/2}$, the
coupling terms $\propto \l_1$ and ${\lambda}_{2}$ will induce
$(x-x_{0})$-independent effects.

Note in particular that, if $\abs{x_{0}}$ is large, as will be the
case when $\Sigma \ll \Lambda$, the dressed $L$ can be much larger than
${\lambda}_{0}$. This implies that, in this case, the nonlinear
couplings can reduce the overall equilibration time {\em both} through
the introduction of a new term $\propto \l_{2}\VEV{E^{2}}$ {\em and}
through an increase in the effective linear coupling.

In any event, to the extent that ${\cal U}(x_{0},{\eta})$ is only slowly
varying in time, the derivative
${{\partial}\VEV{E({\eta})}/{\partial}{\eta}}$ can be replaced by
${{\partial}\VEV{{\cal E}({\eta})}/{\partial}{\eta}}$. And, to the
extent that the coefficients $L$ and ${\lambda}_{2}$ may be
approximated as time-independent, (\ref{63}) can be solved
analytically. The result is an expression identical to (\ref{56}),
except that $E$ is replaced by the shifted ${\cal E}=E-{\cal U}_{0}$
and
\bq
A={L{\omega}_{0}^{~2}\over{\lambda}_{2}k_{B}T}. \label{66}
\eq

If the energy $E$ and the couplings ${\lambda}_{m}$ cannot be treated
as independent of time, the analysis becomes more complicated, but, at
least when $E$ is small, one can again formulate an analogue of
(\ref{58}). If $E$ depends explicitly on ${\eta}$, the moment equation
will of course acquire an additional term
$\VEV{{\partial}E/{\partial}{\eta}}$. Suppose that the system may be
approximated as simply oscillating with squared frequency
${\Sigma}({\eta})$ about $x_{0}=0$. One then concludes that
\bq
{{\partial}E\over {\partial}{\eta}}={1\over 2}\VEV{{\Sigma}x^{2}}
{d\over d\eta}{\rm ln}{\Sigma}. \label{67}
\eq
Consistent, however, with the {\em Ansatz} (\ref{52}),
${\Sigma}x^{2}/2$ can be replaced by its ``orbit averaged'' value
$E/2$ (this is the expectation value associated with the $W$ of eq.
(\ref{52})), so that the moment equation may be written in the form
\bq
{\pa\over\pa\eta}\VEV{E({\eta})} =
{1\over 2}\VEV{E({\eta})}{d\over d\eta}{\rm ln}{\Sigma} + k_{B}T{\lambda}_{0}-
\left({\lambda}_{0}-{k_{B}T{\lambda}_{2}\over
{\Sigma}}\right)\VEV{E({\eta})}- {{\lambda}_{2}\over {\Sigma}}
\VEV{E^{2}({\eta})}. \label{68}
\eq

Suppose now that the time dependence of ${\Sigma}$ is relatively
unimportant, i.e., that the time scale on which ${\Sigma}$ changes is
long compared with the time scale on which the environment effects the
system. In this case, it makes sense to speak of (at least) an
approximate approach towards equilibrium on a time scale set by the
time-dependent couplings. If the coupling between the system and the
environment is dominated by ${\lambda}_{0}$, one thus infers a decay
of initial conditions and an approach towards an equilibrium, driven
by the additive noise, proceeding as ${\rm exp}[-\int
d{\eta}/{\lambda}_{0}({\eta})]$. And similarly, if the coupling is
dominated by ${\lambda}_{2}$, one has an approach towards equilibrium,
driven by the multiplicative noise, proceeding as ${\rm exp}\{-\int
d{\eta}[{\Sigma}/{\lambda}_{2}(k_{B}T{\eta})]\}$.

Suppose, however, that the time dependence of ${\Sigma}$ {\em is}
important, and that ${\Sigma}$ changes appreciably on time scales
$\ll t_{R}$. In this case, one can no longer speak of a simple
approach towards equilibrium, since the form of the system Hamiltonian
is actually changing on a time scale $\ll t_{R}$. One now concludes
that, in a first approximation,
$E({\eta})/E(0)=[{\Sigma}({\eta})/{\Sigma}(0)]^{1/2}$, and that the
coupling with the environment is only a perturbation on this simple
power law evolution.
\newpage

\centerline{\bf VI. Conclusion}
\vskip .1in
The principal thrusts of this paper have been (a) the motivation of a
relatively general phenomenological Hamiltonian, in terms of which
to characterize the evolution of a ``system'' degree of freedom with its
surrounding environment; (b) the rigorous derivation of Langevin and
Fokker-Planck equations for a system described by this model Hamiltonian;
and (c) a qualitative analysis of the new effects incorporated in these
equations which are absent from other, more heuristic, descriptions.
The model considered in this paper has several potential applications,
the most obvious being to the study of cosmological phase transitions.
Unfortunately, however, the Langevin and Fokker-Planck equations
derived here are, except in a few simple cases, very difficult to
solve analytically. For this reason, a numerical study applying the
model to several cosmological problems of interest, such as the onset
of new inflation and first order cosmological phase transitions, is
currently underway.
\vskip 1cm
\centerline{\bf Acknowledgments}
\vskip .1in
S.H. thanks Robert Zwanzig for inspiring discussions and Andreas
Albrecht and Hume Feldman for clarifying remarks regarding their
work. Support for this research was provided by the Department of
Energy and the Air Force Office for Scientific Research. H.E.K. was
supported by the National Science Foundation grant PHY90-03262. Work
on this paper was completed while the authors were visitors at The
Aspen Center for Physics, the hospitality of which is acknowledged
gratefully.

\end{document}